\documentclass[
aps,prd,
12pt,%10pt
nopreprintnumbers,
showpacs,
eqsecnum,
nofootinbib
]{revtex4-1}

\usepackage{graphicx}
\usepackage{amssymb}

\begin{document}

\title{Eisenhart-Duval lift for minisuperspace quantum cosmology}
\author{Nahomi Kan}\email[]{kan@gifu-nct.ac.jp}
\affiliation{National Institute of Technology, Gifu College,
Motosu-shi, Gifu 501-0495, Japan}
\author{Takuma Aoyama}\email[]{b014vbv@yamaguchi-u.ac.jp}
\affiliation{
Graduate School of Sciences and Technology for Innovation, Yamaguchi
University, Yamaguchi-shi, Yamaguchi 753--8512, Japan}
\author{Taiga Hasegawa}\email[]{a019vbu@yamaguchi-u.ac.jp}
\affiliation{
Graduate School of Sciences and Technology for Innovation, Yamaguchi
University, Yamaguchi-shi, Yamaguchi 753--8512, Japan}
\author{Kiyoshi Shiraishi}\email[]{shiraish@yamaguchi-u.ac.jp}
\affiliation{
Graduate School of Sciences and Technology for Innovation, Yamaguchi
University, Yamaguchi-shi, Yamaguchi 753--8512, Japan}
%\author{Zhenyuan Wu}\email[]{b501wb@yamaguchi-u.ac.jp}
\date{\today}
%\date{}

\begin{abstract}
We study covariant equations in quantum cosmology of an extended minisuperspace
obtained by the Eisenhart-Duval lift. We find that a Dirac-type equation is
naturally introduced in the extended minisuperspace. Explicit forms of the
fundamental solutions are yielded for specific models. The possible further
development in this direction is also discussed.
\end{abstract}

%\preprint{}

\pacs{%
%02.10.Ox, %%%Combinatorics; graph theory
%02.20.Sv, %Lie algebra of Lie groups
%02.30.Hq, %Ordinary differential equations
%02.30.Ik, %Integrable systems
%02.30.Jr, %Partial differential equations
%02.40.Gh, %Noncommutative geometry
%03.65.-w, %Quantum mechanics
%03.65.Db, %Functional analytical methods
%03.65.Sq, %Semiclassical theories in quantum mechanics
%03.70.+k, %Theory of quantized fields
%04.20.-q, %%%Classical general relativity
%04.20.Fy, %%Canonical formalism, Lagrangians, and variational principles
%04.20.Jb, %%Exact solutions
%04.25.-g, %Approximation
%04.25.Nx, %%%Post-Newtonian approximation; perturbation theory; related
%approximations
%04.40.-b, %Self-Gravitating systems
%04.40.Nr, %%Einstein-Maxwell spacetime
%04.50.-h, %%%%%Higher-dimensional gravity and other theories of gravity 
%04.50.Cd, %Kaluza-Klein theories 
%04.50.Gh, %Higher-dimensional black holes, black strings, 
%and related objects 
%04.50.Kd, %%%Modified theories of gravity 
04.60.-m, %%Quantum gravity
%04.60.Kz, %%Lower dimensional models; minisuperspace models
%04.60.Rt, %Topologically massive gravity
%04.62.+v, %Quantum fields in curved spacetime
%04.65.+e, %Supergravity
%04.70.Bw, %%%Classical black holes
%05.30.Jp, %Boson systems
%11.10.-z, %%%Field theory
%11.10.Lm, %%%Nonlinear or nonlocal theories and models 
%11.10.Nx, %%%Noncommutative field theory 
%11.10.Kk, %%%Field theories in dimensions other than four
%11.25.-w, %Strings and branes
%11.25.Mj, %%Compactification and four-dimensional models
%11.27.+d% %%Extended classical solutions; cosmic strings, 
%domain walls, texture 
11.30.-j, %Symmetry and conservation laws
%11.30.Pb, %Supersymmetry
%12.60.-i, %Models beyond the standard model
45.20.Jj, %Lagrangian and Hamiltonian mechanics
%95.35.+d, %Dark matter
%95.36.+x, %Dark energy
%98.80.-k, %%%Cosmology 
%98.80.Cq, %%%%%Particle-theory and field-theory models of the early
%Universe  
%98.80.Dr, %Relativistic cosmology 
98.80.Qc, %Quantum cosmology
98.80.Jk% %%Mathematical and relativistic aspects of cosmology
.}

\maketitle

%%%%%%%%%%%%%%%%%%%%%%%%%%%%%%%%%%%%%%%%%%%%%%%%%%%%%%%%%%%%%%%%%%%%%%%%%%%
%%%%%%%%%%%%%%%%%%%%%%%%%%%%%%%%%%%%%%%%%%%%%%%%%%%%%%%%%%%%%%%%%%%%%%%%%%%
%%%%%%%%%%%%%%%%%%%%%%%%%%%%%%%%%%%%%%%%%%%%%%%%%%%%%%%%%%%%%%%%%%%%%%%%%%%
\section{Introduction}
\label{introduction}
%%%%%%%%%%%%%%%%%%%%%%%%%%%%%%%%%%%%%%%%%%%%%%%%%%%%%%%%%%%%%%%%%%%%%%%%%%%
%%%%%%%%%%%%%%%%%%%%%%%%%%%%%%%%%%%%%%%%%%%%%%%%%%%%%%%%%%%%%%%%%%%%%%%%%%%
%%%%%%%%%%%%%%%%%%%%%%%%%%%%%%%%%%%%%%%%%%%%%%%%%%%%%%%%%%%%%%%%%%%%%%%%%%%

Quantum cosmology \cite{HH,Hawking,Halliwell,Kiefer0,Kiefer1} has some problems to
be solved. One of them is the process of deriving the probability density from
the wave function of the universe. Since the usual Wheeler-DeWitt (WDW) equation
in minisuperspace is a hyperbolic partial differential equation, it is very
difficult to define a positive-definite probability density.
Note that a component of the Klein-Gordon type conserved current
may vanish if the wave function is substantially expressed by a real function,
which often occurs as a solution of the WDW equation.
Several hopeful approaches have been proposed to replace the WDW equation with
other equations. For example, many authors have already proposed ideas to
attempt a square root \`a la Dirac of the quadratic WDW equation
\cite{DHO,KO,SC,YH,HA,RAH}, using two component wave functions
\cite{Mostafazadeh}, and
applications of supersymmetric quantum mechanics
\cite{Graham1,Graham2,OSB,OPR,RB},%
\footnote{For a review of supersymmetric quantum mechanics, please see
Ref.\cite{CKS}.}
 and the introduction of new
``time'' coordinates \cite{Hajicek,Simeone,CF,Kuzmichev,KKST}.

Another problem with quantum cosmology is the problem of factor ordering%
\footnote{Although the general factor-ordering problem in quantum gravity is known
to be more difficult than the arrangement of momentum operators (see, for
example, Ref.~\cite{TW}) our present analysis is
limited to the factor ordering on momentum operators.}   which arises with
substitutions of operators for momenta
\cite{HP,Moss,Halliwell2}. In some toy models, a wave function with little
difference can be obtained regardless of the choice in factor ordering of
momentum operators to some extent.  However, the choice is advocated to be
important in some cases when considering various boundary conditions \cite{KW}.
In particular, applying the Dirac square-root method mentioned above, it is also
a considerable problem that the choice of ordering involves additional
indefiniteness. In any case, when we wish to discuss some hidden mathematical
structures and symmetry of gravitational theory minutely, we should not neglect
the problem of factor ordering altogether.

Recently, remarkable papers \cite{FK,Finn} have appeared that apply a type of the
Eisenhart-Duval lift \cite{Eisenhart,Duval}, one of the classical methods in
Hamiltonian dynamical systems \cite{CPC,Pettini,Cariglia}%
\footnote{See also a recent paper, Ref.~\cite{DSS}.} to cosmologies.%
\footnote{See also Ref.~\cite{CGGH,DGH}.} In this Eisenhart's method, adding a
dynamical variable, it is possible to describe a system by geometric
treatment in the space of dynamical variables; that is, in the extended
minisuperspace even in the presence of the potential term. Thus, the Hamiltonian
of the system can be represented by the single Laplacian on the extended
minisuperspace when the momenta are replaced by operators.
%RR
It should be noted that the covariance in the minisuperspace is not the
general covariance of spacetime. The original idea of Eisenhart's work is to
interpret a generic (nongeometrical) equation of motion as a (geometrical)
geodesic equation in a space with a lifted metric.
%RR

We come to an idea that, if
covariance in the extended minisuperspace is required as a guiding principle, the
problem of factor ordering in the WDW equation disappears and it becomes possible
to proceed with covariance as a prescription for Dirac square root. 

In the present paper we consider the construction of covariant equations using
the Eisenhart-Duval lift in minisuperspace quantum cosmology. For the
sake of simplicity, we will focus on the case of a homogeneous and isotropic 
Friedmann-Lema\^{\i}tre-Robertson-Walker (FLRW)
universe containing a single spatially constant scalar field (that is, the case of
two dynamical variables) in this paper. Section \ref{sec2} discusses the
extension of the minisuperspace by the Eisenhart-Duval lift and possible
conformal invariance of the WDW equations. We study the treatment of additional
degrees of freedom especially in the known simple models. In Sec.~\ref{sec3}, we
propose quantum cosmology with the Dirac equation in the extended minisuperspace
of the simple models, and their fundamental solutions are presented. The last
section will be devoted to discussions and future prospects.

%%%%%%%%%%%%%%%%%%%%%%%%%%%%%%%%%%%%%%%%%%%%%%%%%%%%%%%%%%%%%%%%%%%%%%%%%%%
%%%%%%%%%%%%%%%%%%%%%%%%%%%%%%%%%%%%%%%%%%%%%%%%%%%%%%%%%%%%%%%%%%%%%%%%%%%
%%%%%%%%%%%%%%%%%%%%%%%%%%%%%%%%%%%%%%%%%%%%%%%%%%%%%%%%%%%%%%%%%%%%%%%%%%%
\section{Minisuperspace extended by the Eisenhart-Duval lift and geometry of the
WDW equation}
\label{sec2}
%%%%%%%%%%%%%%%%%%%%%%%%%%%%%%%%%%%%%%%%%%%%%%%%%%%%%%%%%%%%%%%%%%%%%%%%%%%
%%%%%%%%%%%%%%%%%%%%%%%%%%%%%%%%%%%%%%%%%%%%%%%%%%%%%%%%%%%%%%%%%%%%%%%%%%%
%%%%%%%%%%%%%%%%%%%%%%%%%%%%%%%%%%%%%%%%%%%%%%%%%%%%%%%%%%%%%%%%%%%%%%%%%%%

In this paper we consider a homogeneous and isotropic universe containing a
single spatially constant scalar field. We will just add comments for generic
cases occasionally. First of all, we would like to review the ``conventional''
WDW equation, and show our approach with the Eisenhart-Duval lift afterwards.

At first, we start with the following action of the
gravitating scalar field
\begin{equation}
S=\int d^4x\sqrt{-g}\left[\frac{1}{2\kappa^2}R-\frac{1}{2}(\nabla\phi)^2
-V(\phi)
\right]\,,
\label{action1}
\end{equation}
where $g$ is the determinant of the metric tensor
$g_{\mu\nu}$,
$R$ denotes the scalar curvature constructed from $g_{\mu\nu}$ $(\mu,\nu=0,1,2,3)$,
$(\nabla\phi)^2$ means
$g^{\mu\nu}
\partial_\mu\phi\partial_\nu\phi$, and $V(\phi)$ represents a potential for the
real scalar field $\phi$. The constant $\kappa^2$ equals to $8\pi G$, where $G$ is
Newton's constant.  

As the metric, we assume the FLRW metric,
\begin{equation}
ds^2=-N^2dt^2+a^2(t)d\Omega_3^2\,,
\end{equation}
where $d\Omega_3^2$ denotes the maximally symmetric three-space,
whose Ricci curvature ${}^{(3)}R_{ij}$ is characterized by a constant $K$, such as
${}^{(3)}R_{ij}=2Kg_{ij}$ $(i,j=1,2,3)$, while $N$ is the lapse function. Assuming
that the scalar field depends only on time
$t$, the Lagrangian can be written in the following form%
\footnote{Here we have added the standard Gibbons-Hawking-York boundary term
\cite{York,GH}.}
\begin{equation}
L=-\frac{1}{2N}a\dot{a}^2
+\frac{1}{2N}a^3\dot{\phi}^2-NU(a,\phi)\,,
\end{equation}
where the dot denotes the time derivative, and the potential term $U(a,\phi)$ is,
for the above action (\ref{action1}),
\begin{equation}
U(a,\phi)=a^3V(\phi)-\frac{1}{2}Ka\,.
\end{equation}
In the derivation of the Lagrangian the physical units are chosen to be
$\kappa^2=6$, for convenience.

The canonical analysis of this action defines the Hamiltonian of the system.
The lapse function $N$ plays the role of a Lagrange multiplier and we find that the
Hamiltonian constraint condition $H=0$, where
\begin{equation}
H=-\frac{1}{2}\frac{\Pi_a^2}{a}
+\frac{1}{2}\frac{\Pi_\phi^2}{a^3}+U(a,\phi)\,,
\end{equation}
with the conjugate momenta to the scale
factor and the scalar field, $\Pi_a=-\frac{a\dot{a}}{N}$ and
$\Pi_\phi=\frac{a^3\dot{\phi}}{N}$, respectively.

In quantum gravity the Hamiltonian constraint acting on states leads to a
differential equation, known as the WDW equation. We interpret that the solution
of the WDW equation is the physical states. Preparing a wave function $\Psi$ as a
state and replacing the momenta with differential operators as
\begin{equation}
\Pi_a\rightarrow -i\frac{\partial}{\partial a}\,,\quad
\Pi_\phi\rightarrow -i\frac{\partial}{\partial \phi}\,,
\end{equation}
we find the usual WDW equation in minisuperspace quantum cosmology
\cite{HH,Hawking,Halliwell,Kiefer0,Kiefer1}
\begin{equation}
\left[\frac{1}{a^{s+1}}\frac{\partial}{\partial a}a^s\frac{\partial}{\partial
a}-\frac{1}{a^3}\frac{\partial^2}{\partial
\phi^2}+2 U(a,\phi)\right]\Psi(a,\phi)=0\,,
\label{convWDW}
\end{equation}
where the constant $s$ indicates the arbitrariness in factor ordering.
So far, we have obtained a description for the derivation of the
conventional WDW equation.

 Now, let us consider an extension with a new degree of freedom. This has been
proposed in previous studies \cite{FK,Finn} as a type of the Eisenhart-Duval
lift. In the present case, the lifted Lagrangian (which includes additional
variable $\chi$) is given by%
\footnote{The lapse function $N$ is omitted here, since the procedure is
straightforward around this step.}
\begin{equation}
\tilde{L}=-\frac{1}{2}a\dot{a}^2
+\frac{1}{2}a^3\dot{\phi}^2+\frac{1}{2}\frac{\dot{\chi}^2}{2U(a,\phi)}
=\frac{1}{2}\tilde{G}_{MN}\dot{X}^M\dot{X}^N\,,
\end{equation}
where $X^M=(a,\phi, \chi)$, and 
the metric of the extended minisuperspace is 
\begin{equation}
\tilde{G}_{MN}=\mbox{diag} (-a, a^3, [2U(a,\phi)]^{-1})\,.
\end{equation}

Using this extended metric the Hamiltonian constraint of the system is simply
written as
\begin{equation}
\frac{1}{2}\tilde{G}^{MN}\tilde{P}_M\tilde{P}_N=0\,,
\label{classical}
\end{equation}
where the momenta $\tilde{P}_M$ is given by $\tilde{P}_M=\tilde{G}_{MN}\dot{X}^N$
and $\tilde{G}^{MN}$ denotes the inverse of the metric $\tilde{G}_{MN}$ as usual.
This classical equation (\ref{classical}) has an apparent classical conformal
invariance under $\tilde{G}_{MN}\rightarrow \Omega^2 \tilde{G}_{MN}$
with an arbitrary function $\Omega(X^M)$. Here
we would like to propose the following choice for extended
three-dimensional minisuperspace quantum cosmology, at least because it simplifies
the equation. We take the following ``gauge choice'', or in practice, we use the
following metric of the extended minisuperspace%
\footnote{In the case of $n$ dimensional extended minisuperspace, we should
take $G_{MN}=[2U(a,\phi)]^{\frac{1}{n-2}}\tilde{G}_{MN}$.}
\begin{equation}
G_{MN}=2U(a,\phi)\tilde{G}_{MN}=\mbox{diag} (-2U(a,\phi)a, 2U(a,\phi)a^3,
1)\,.
\label{gauge}
\end{equation}

There is factor-ordering ambiguity in promoting the conjugate variables to
 differential operators, $P_M\sim -i\frac{\partial}{\partial X^M}$ in this
case. A possible choice is to use the Laplace-Beltrami operator as the quadratic
operator. In addition, we can add a term proportional to the scalar curvature
${\cal R}$ of the extended minisuperspace, which can appear in quantum mechanical
systems.%
\footnote{Note that the dimension of $P^2$ equals to that of $\hbar^2{\cal R}$.}
Consequently, the ``extended'' WDW equation is written by the Klein-Gordon
equation in the extended minisuperspace as
\begin{equation}
\left[\frac{1}{\sqrt{-{G}}}\partial_M\sqrt{-{G}}{G}^{MN}
\partial_N-\xi{\cal R}\right]\Psi=0\,,
\label{scon}
\end{equation}
which has covariance in the extended minisuperspace.
Here, $G^{MN}$ is the inverse of $G_{MN}$, $G=-(2U)^2a^4$ is the determinant of
$G_{MN}$, the derivatives are expressed as
$\partial_M\equiv\frac{\partial}{\partial X^M}$, and
$\xi$ is a dimensionless constant.

Here, ${\cal R}$ is the scalar curvature constructed from $G_{MN}$,
which is defined by
\begin{equation}
{\cal
R}=G^{MN}\left(\partial_{L}\Gamma^{L}_{MN}-\partial_{M}\Gamma^{L}_{NL}
+\Gamma^{L}_{MN}\Gamma^{P}_{LP}-\Gamma^{L}_{MP}\Gamma^{P}_{NL}\right)\,,
\end{equation}
where the Christoffel symbol $\Gamma^L_{MN}$ is given by
\begin{equation}
\Gamma^L_{MN}=\frac{1}{2}G^{LP}(\partial_MG_{PN}+\partial_NG_{PM}
-\partial_PG_{MN})\,.
\end{equation}

The explicit form of the extended WDW equation is written by
\begin{eqnarray}
& &\left\{\frac{1}{a^{2}}\frac{\partial}{\partial a}a\frac{\partial}{\partial
a}-\frac{1}{a^3}\frac{\partial^2}{\partial
\phi^2}\right.\nonumber \\
& &\left.-\left(2a^3V-Ka\right)\frac{\partial^2}{\partial\chi^2}
+2\xi\frac{2a^3[(V')^2-V''V]+Ka[V''-4
V]}{\left(2a^3V(\phi)-Ka\right)^2}\right\}\Psi(a,\phi,\chi)=0\,,
\label{hc}
\end{eqnarray}
where $V'=\frac{\partial V}{\partial\phi}$ and $V''=\frac{\partial^2
V}{\partial\phi^2}$.

If we require another constraint on a physical state
\begin{equation}
-\frac{\partial^2}{\partial \chi^2}\Psi=p^2\Psi\,,
\label{P}
\end{equation}
the conventional WDW equation (\ref{convWDW}) is recovered provided that we set
the constants $p^2=1$ and
$\xi=0$. Although it sounds unhealthy
to add constraints without permission, it should be noted that the WDW
wave functions have been introduced intentionally for the purpose of expressing
the Hamiltonian constraint in the first place.%R
\footnote{The two constraints (\ref{hc}) and (\ref{P}) are compatible because
they form a first-class system of constraints.} The specific value of $p^2$ is
actually not significant because the scaling by a constant can be absorbed in
the primary definition of the time coordinate in the FLRW metric of our universe.
The interpretation of the value of $p$ will be further discussed later.

The classical equation (\ref{classical}) has conformal
symmetry. Therefore, it is natural to use conformal Laplacian (Yamabe-Laplacian)
when making the quantum equation. This adoption has also been suggested in the
conventional minisuperspace quantum cosmology
\cite{Moss,Halliwell2}.
%\footnote{Note that there is no conformal symmetry in general, because
%the extra potential term is generally added in the conventional WDW equation.}
Since the extended minisuperspace in the present model is three dimensional, the
conformal coupling is $\xi=\frac{1}{8}$, while  $\xi=\frac{n-2}{4(n-1)}$ in the
case of $n$-dimensional extended minisuperspaces.

For a nonzero value of $\xi$, the WDW equation may become a different form the
conventional one since the contribution from the curvature term ${\cal R}$ is
added in general.%
\footnote{Note that the curvature term ${\cal R}$ becomes small if $K\approx 0$ and
the slow-roll parameters \cite{PDG} of the potential are small (i.e.,
$\left(\frac{V'}{V}\right)^2\ll 1$ and $\frac{V''}{V}\ll 1$) as an inflaton
potential.} The ``nice'' value for
$\xi$ has been discussed for quite a long time in the field of quantum field
theory in curved spacetime
\cite{Parker,BP}. The value might depend on the functional measure in the path
integral on manifolds
\cite{Laetsch} and we feel that ambiguity is also left despite numerous
discussions with interpretation of the WDW wave function. Incidentally, the
continuum limit of the d'Alembertian acting on scalar fields in causal set theory
predicts the other value of the scalar curvature coupling
\cite{BD,DG,Glaser,Belenchia,BBD}. 
It is also known that in the stochastic quantization of geodesics, a coupling
term of a scalar field and the scalar curvature emerges regardless of the
presence or absence of conformal symmetry \cite{Kuipers1,Kuipers2}. 
 There must still be many issues to be
investigated on the coupling $\xi$ even in quantum cosmology in detail. 
The correction term of the WDW equation due to
${\cal R}$ itself needs to be examined more carefully for its properties, which is
an important topic for the future.

There exists an interesting class of extended minisuperspace models for which the
scalar curvature ${\cal R}$ vanishes. In such models, the extended WDW equation
(\ref{scon}) with the constraint (\ref{P}) coincides with the conventional one,
regardless of the value of $\xi$ (in the case of the present gauge choice).
In the present Einstein minimally-coupled scalar system, ${\cal R}=0$ is satisfied
in the following two cases: 1) $V(\phi)\equiv 0$, i.e., free scalar case, and 2)
$K=0$ and
$V(\phi)=V_0 \exp(\lambda\phi)$, where $V_0$ and $\lambda$ are constants, i.e.,
the case with a Liouville-type potential in a flat universe. It is interesting to
point out that the WDW equation in each model is known to be analytically solvable
\cite{HP2,Kiefer0,Kiefer2,ALNW}.%
\footnote{The author of Ref.~\cite{Paliathanasis} solved the WDW equation for
the case of exponential scalar field with the dust matter.}

%%%%%%%%%%%%%%%%%%%%%%%%%%%%%%%%%%%%%%%%%%%%%%%%
\if0
%%%%%%%%%%%%%%%%%%%%%%%%%%%%%%%%%%%%%%%%%%%%%%%%
For the system with a conformally coupled scalar field (without
potential) described by the following actions, 
\begin{equation}
S=\int d^4x\sqrt{-g}\left[\frac{1}{12}R-\frac{1}{2}(\nabla\phi)^2
-\frac{1}{12} R\phi^2\right]\,,
\end{equation}
the minisuperspace extended by the Eisenhart-Duval lift has the metric
\begin{equation}
G_{MN}=\mbox{diag} (-2U(a,\Phi)a, 2U(a,\Phi)a,
1)\,,
\end{equation}
with $U(a,\Phi)=\frac{Ka}{2}\left(\frac{\Phi^2}{2}-1\right)$,
taking the variable $\Phi=a\phi$ instead of $\phi$, so $X^M=(a,\Phi,\chi)$. 
Then, the curvature of the extended minisuperspace is also zero.%
\footnote{Actually, even in the case of a general non-minimal coupling
constant, the extended minisuperspace is found to be flat. The extended
minisuperspace for the Wudka model \cite{Wudka1,Wudka2}, which is related to the
similar harmonic oscillator, is also flat.}
%%%%%%%%%%%%%%%%%%%%%%%%%%%%%%%%%%%%%%%%%%%%%%%%
\fi
%%%%%%%%%%%%%%%%%%%%%%%%%%%%%%%%%%%%%%%%%%%%%%%%

The WDW equations for the two models, of which extended minisuperspace is flat,
are explicitly written as

\noindent$(i)$ \underline{Model 1} (the case of $V=0$):
\begin{equation}
\left[a\frac{\partial}{\partial a}a\frac{\partial}{\partial
a}-\frac{\partial^2}{\partial
\phi^2}+Ka^4\frac{\partial^2}{\partial\chi^2}
\right]\Psi(a,\phi,\chi)=0\,,
\label{KGE1}
\end{equation}

\noindent$(ii)$ \underline{Model 2} (the case of $K=0$ and
$V(\phi)=V_0\exp\lambda\phi$):
\begin{equation}
\left[a\frac{\partial}{\partial a}a\frac{\partial}{\partial
a}-\frac{\partial^2}{\partial
\phi^2}-2a^6V_0\exp\lambda\phi\frac{\partial^2}{\partial\chi^2}
\right]\Psi(a,\phi,\chi)=0\,.
\label{LEKG2}
\end{equation}

%%%%%%%%%%%%%%%%%%%%%%%%%%%%%%%%%%%%%%%%%%%%%%%%
\if0
%%%%%%%%%%%%%%%%%%%%%%%%%%%%%%%%%%%%%%%%%%%%%%%%
\noindent$\bullet$ the case of conformal scalar field without the scalar potential
(Model 3):
\begin{equation}
\left[\frac{\partial^2}{\partial
a^2}-\frac{\partial^2}{\partial
\Phi^2}-K\left(\Phi^2-a^2\right)\frac{\partial^2}{\partial\chi^2}
\right]\Psi(a,\Phi,\chi)=0\,.
\end{equation}
%%%%%%%%%%%%%%%%%%%%%%%%%%%%%%%%%%%%%%%%%%%%%%%%
\fi
%%%%%%%%%%%%%%%%%%%%%%%%%%%%%%%%%%%%%%%%%%%%%%%%

If we additionally assume 
\begin{equation}
-\frac{\partial^2}{\partial \chi^2}\Psi=p^2\Psi\,,
\end{equation}
we can see that each wave function satisfies the conventional WDW equation in each
case, by performing the following redefinition of variables in each model
\begin{eqnarray}
\mbox{\underline{Model 1}:}&\qquad&\sqrt{|p|}a\rightarrow a \,,\\
\mbox{\underline{Model 2}:}&\qquad&\beta a\rightarrow a \mbox{ and }
\phi+\gamma\rightarrow
\phi\,, \mbox{ where }\beta^6e^{\lambda\gamma}=p^2 \,.
\end{eqnarray}
\if0
\begin{equation}
\sqrt{|p|}a\rightarrow a \mbox{ and }  \sqrt{|p|}\Phi\rightarrow \Phi\mbox{
(Model 3)}\,.
\end{equation}
\fi
Therefore, in these cases, the setting $p^2=1$ considered first is not a special
one. Moreover, from the result we can obtain solutions of the extended WDW
equation (even if the variable $\chi$ is not a fictitious one)
as the superposition of the solutions of the conventional WDW equation
under different boundary conditions.
To interpret the extended WDW equation (not as a mathematical manipulation but as
a fundamental one) is very interesting and may be an approach to the initial
condition problem of the Universe. Note, however, that the models we found so far
have no tunneling potential and we are restricted to take the wave-packet
interpretation \cite{KN,Kiefer2,Kiefer3,ALNW,GS} in these special cases.
 The analysis of the more general case will be left as a future task.

Before closing this section, we put the forms of solutions in Models 1--2
below, which will be compared with the result of the Dirac-like first-order
differential equation in the extended minisuperspace introduced in the next
section. For the cases, general solutions are given by superposition of  the
fundamental solutions
${\cal A}(\nu)\psi_{\nu p}e^{ip\chi}$, where ${\cal A}(\nu)$
$(-\infty<\nu<\infty)$ is an appropriate amplitude.

\noindent$(i)$ \underline{Model 1}:\cite{HP2,Kiefer0,Kiefer2}%R
\footnote{For the case of $K=0$ is rather trivial (and includes no effective
potential), so we do not deal with the case in this paper.}
\begin{equation}
\psi_{\nu p}=K_{
i\nu/2}(\sqrt{K}|p|a^2/2)e^{i\nu(\phi-\phi_0)}\qquad (K>0)\,,
\label{KGS1}
\end{equation}
\begin{equation}
\psi_{\nu p}=J_{\pm
i\nu/2}(\sqrt{|K|}|p|a^2/2)e^{i\nu(\phi-\phi_0)}\qquad (K<0)\,,
\end{equation}
where $\phi_0$ is a constant, and the functions $J_\nu(z)$ and $K_\nu(z)$ are the
Bessel function and the modified Bessel function of the second kind, respectively.

\noindent$(ii)$ \underline{Model 2}:\cite{ALNW}
\begin{equation}
\psi_{\nu p}=J_{\pm
i\nu/3}(i\sqrt{C}|p|e^{3x}/3)e^{i\nu(y-y_0)}\qquad (C>0)\,,
\end{equation}
\begin{equation}
\psi_{\nu p}=K_{
i\nu/3}(i\sqrt{|C|}|p|e^{3x}/3)e^{i\nu(y-y_0)}\qquad (C<0)\,,
\end{equation}
where $x\equiv\alpha+\frac{\lambda}{6}\phi$,
$y\equiv\phi+\frac{\lambda}{6}\alpha$ with $\alpha=\ln a$, and $y_0$ is a
constant. The constant $C$ represents
$2V_0\left(1-\frac{\lambda^2}{36}\right)^{-1}$.

\if0
\noindent$\bullet$ the case of conformal scalar field without the scalar potential
(Model 3):
\begin{equation}
\psi_{\nu p}=\sqrt{r}K_{
\sqrt{1+4\nu^2}/4}(\sqrt{K}|p|r^2/2)e^{i\nu \theta}\qquad (K>0)\,,
\end{equation}
\begin{equation}
\psi_{\nu p}=\sqrt{r}J_{\pm
\sqrt{1+4\nu^2}/4}(\sqrt{|K|}|p|r^2/2)e^{i\nu \theta}\qquad (K<0)\,,
\end{equation}
where $r$ and $\theta$ are defined through
$a=r\cosh\theta$ and $\Phi=r\sinh\theta$.
\fi

%%%%%%%%%%%%%%%%%%%%%%%%%%%%%%%%%%%%%%%%%%%%%%%%%%%%%%%%%%%%%%%%%%%%%%%%%%%
%%%%%%%%%%%%%%%%%%%%%%%%%%%%%%%%%%%%%%%%%%%%%%%%%%%%%%%%%%%%%%%%%%%%%%%%%%%
%%%%%%%%%%%%%%%%%%%%%%%%%%%%%%%%%%%%%%%%%%%%%%%%%%%%%%%%%%%%%%%%%%%%%%%%%%%
\section{Dirac equation in the extended minisuperspace}
\label{sec3}
%%%%%%%%%%%%%%%%%%%%%%%%%%%%%%%%%%%%%%%%%%%%%%%%%%%%%%%%%%%%%%%%%%%%%%%%%%%
%%%%%%%%%%%%%%%%%%%%%%%%%%%%%%%%%%%%%%%%%%%%%%%%%%%%%%%%%%%%%%%%%%%%%%%%%%%
%%%%%%%%%%%%%%%%%%%%%%%%%%%%%%%%%%%%%%%%%%%%%%%%%%%%%%%%%%%%%%%%%%%%%%%%%%%

The idea of taking the square root of the WDW equation can be found in
Refs.~\cite{DHO,KO,SC,YH,HA,RAH} and others.%
\footnote{There is another example of applying supersymmetric
quantum mechanics as another method for deriving first-order differential
equations \cite{Graham1,Graham2,OSB,OPR,RB}.} In previous studies, 
arbitrariness inevitably remains in the treatment of the potential term in
the conventional WDW equation. Our treatment on how to proceed is now clear; to
use the Dirac equation in the extended minisuperspace instead of the WDW equation
of Klein-Gordon-type. In this case, the possible form of the equation is fixed
from the covariance in the extended minisuperspace. Also note that the Dirac
equation (without the mass term) has conformal covariance.%
\footnote{See, for instance, Ref.~\cite{Hijazi}.}
We suppose that the gauge choice is the same as in the previous
section, i.e., we adopt $G_{MN}$ as the metric of the extended
minisuperspace, though there is no guarantee that the choice of the metric will
commonly make the equation easy to solve.

%From the conformal covariance of the Dirac equation,
%we can always obtain the Dirac wave function $\tilde{\Psi}$ in the space with the
%metric $\tilde{G}_{MN}$ as $\tilde{\Psi}=\sqrt{2U}\Psi$,%
%\footnote{If the space has $n$ dimensions and $G_{MN}=\Omega^2\tilde{G}_{MN}$,
%$\tilde{\Psi}=\Omega^{\frac{n-1}{2}}\Psi$.}
%where
%$\Psi$ is the Dirac wave function in the space with the metric $G_{MN}$.

The Dirac-like equation in the extended minisuperspace can be written down as
\begin{equation}
D\!\!\!\!/\,\Psi\equiv\hat{\gamma}^MD_M\Psi\equiv\gamma^A e_A^MD_M\Psi=0\,.
\label{DE}
\end{equation}
Here, the constant gamma matrices in the flat spacetime $\gamma^A$ ($A=1,2,3$) are
$\gamma^1=\sigma^1$, $\gamma^2=i\sigma^2$, and $\gamma^3=i\sigma^3$,
where $\sigma^1$, $\sigma^2$, $\sigma^3$ are the Pauli matrices. Note that
$\{\gamma^A,
\gamma^B\}=-2\eta^{AB}$, where $\eta^{AB}=\eta_{AB}=\mbox{diag}(-1,1,1)$.
The dreibein $e^A_M=\mbox{diag} ((2U)^{1/2}a^{1/2}, (2U)^{1/2}a^{3/2},
1)$ is defined through $\eta_{AB}e^A_Me^B_N=G_{MN}$,
and $e_A^M=\mbox{diag} ((2U)^{-1/2}a^{-1/2}, (2U)^{-1/2}a^{-3/2},
1)$ is its inverse matrix. Accordingly, we find that $\{\hat{\gamma}^M,
\hat{\gamma}^N\}=-2G^{MN}$.

The covariant derivative $D_M$ for the spin connection $\omega_{MAB}$ is defined
as
$D_M\equiv\partial_M+\frac{1}{4}
\omega_{MAB}\Sigma^{AB}$, where $\Sigma^{AB}\equiv-\frac{1}{2}[\gamma^A,
\gamma^B]$. The spin connection $\omega_{MAB}$ is given by
\begin{equation}
\omega_{MAB}=\frac{1}{2}e^N_A(\partial_Me_{NB}-\Gamma^L_{MN}e_{LB})
-(A\leftrightarrow B)\,.
\end{equation}

In this section we shall specifically demonstrate the analysis of the simplest
cases; Models 1 and 2 introduced in the previous section. 

In Model 1 we find that the Dirac equation (\ref{DE}) is equivalent to
\begin{equation}
\left[\sigma^1\left(a\frac{\partial}{\partial a}+1\right)+
i\sigma^2\frac{\partial}{\partial
\phi}+
i\sigma^3\sqrt{-K}a^2\frac{\partial}{\partial
\chi}\right]\Psi=0\,.
\end{equation}
We set the two components of the wave function as
\begin{equation}
\Psi={\Psi_{+}\choose\Psi_{-}}e^{ip\chi}
\end{equation}
in order to find
solutions of it. Now, the equation reads in the matrix form
\begin{equation}
\left(
\begin{array}{cc}
-p\sqrt{-K}a^2 &
a\frac{\partial}{\partial
a}+1+\frac{\partial}{\partial\phi}\\
a\frac{\partial}{\partial
a}+1-\frac{\partial}{\partial\phi} &
p \sqrt{-K}a^2
\end{array}
\right)\left(
\begin{array}{c}
\Psi_{+} \\ \Psi_{-}
\end{array}
\right)=\left(
\begin{array}{c}
0 \\ 0
\end{array}
\right)\,.
\label{DE1}
\end{equation}
The fundamental solutions of the equation are%R
\footnote{Just as in the case of second-order differential equations, there
are generally two independent fundamental solutions, but solutions with regions
where their absolute values are infinite are rejected.\label{fn}}%RR
\begin{equation}
\Psi_{\pm,\nu p}={\frac{1}{\sqrt{2}}}e^{\pm i\frac{\pi}{4}}K_{
\frac{i\nu}{2}\mp\frac{1}{2}}(\sqrt{K}|p|a^2/2)e^{i\nu(\phi-\phi_0)}\quad
(K>0)\,,
\label{DS1}
\end{equation}
\begin{equation}
\Psi_{\pm,\nu
p}={\scriptstyle\frac{1}{\sqrt{2}}}J_{\frac{i\nu}{2}\mp\frac{1}{2}}(\sqrt{|K|}|p|a^2/2)e^{i\nu
(\phi-\phi_0)}\,,~\pm
{\scriptstyle\frac{1}{\sqrt{2}}}J_{-\frac{i\nu}{2}\pm\frac{1}{2}}(\sqrt{|K|}|p|a^2/2)e^{i\nu
(\phi-\phi_0)}~ (K<0)\,,
\end{equation}
where $-\infty<\nu<\infty$.

%%%%%%%

In Model 2 we find that the Dirac equation (\ref{DE}) is equivalent to
\begin{equation}
\left[\sigma^1\left(a\frac{\partial}{\partial a}+\frac{3}{2}\right)+
i\sigma^2\left(\frac{\partial}{\partial
\phi}+\frac{\lambda}{4}\right)+
i\sigma^3\sqrt{2V_0e^{\lambda\phi}}a^3\frac{\partial}{\partial
\chi}\right]\Psi=0\,.
\end{equation}
We set 
$\Psi={\Psi_{+}\choose\Psi_{-}}e^{ip\chi}$ and then the equation
reads in the matrix form, 
\begin{equation}
\left(
\begin{array}{cc}
-pa^3\sqrt{2V_0e^{\lambda\phi}} &
a\frac{\partial}{\partial
a}+\frac{3}{2}+\frac{\partial}{\partial
\phi}+\frac{\lambda}{4}\\
a\frac{\partial}{\partial
a}+\frac{3}{2}-\frac{\partial}{\partial
\phi}-\frac{\lambda}{4} &
p a^3\sqrt{2V_0e^{\lambda\phi}}
\end{array}
\right)\left(
\begin{array}{c}
\Psi_{+} \\ \Psi_{-}
\end{array}
\right)=\left(
\begin{array}{c}
0 \\ 0
\end{array}
\right)\,.
\label{LE2}
\end{equation}
The fundamental solutions of the equation are%
\footnote{Please see footnote~\ref{fn}.} %RR
\begin{equation}
\Psi_{\pm,\nu
p}={\scriptstyle\frac{1}{\sqrt{2}}}{\scriptstyle\sqrt{1\pm\frac{\lambda}{6}}}
J_{\frac{i\nu}{3}\mp\frac{1}{2}}(\sqrt{C}|p|e^{3x}/3)e^{i\nu (y-y_0)}\,,~~\pm
{\scriptstyle\frac{1}{\sqrt{2}}}{\scriptstyle\sqrt{1\pm\frac{\lambda}{6}}}J_{-\frac{i\nu}{3}\pm\frac{1}{2}}
(\sqrt{C}|p|e^{3x}/3)e^{i\nu
(y-y_0)}\quad (C>0)\,.
\end{equation}
\begin{equation}
\Psi_{\pm,\nu p}={\scriptstyle\frac{1}{\sqrt{2}}}e^{\pm
i\frac{\pi}{4}}{\scriptstyle\sqrt{1\pm\frac{\lambda}{6}}}K_{
\frac{i\nu}{3}\mp\frac{1}{2}}(\sqrt{|C|}|p|e^{3x}/3)e^{i\nu(y-y_0)}\qquad (C<0)\,,
\end{equation}
where the definition of $x$ and $y$ is the same as in the previous section.

Probability density is endowed by the square of the norm of the wave function
\cite{DHO,KO,SC,YH,HA,RAH}
\begin{equation}
\mbox{Probability density}\propto
\sqrt{|2U|}a^{3/2}\|\Psi\|^2=\sqrt{|2U|}a^{3/2}(|\Psi_+|^2+|\Psi_-|^2)\,,
\end{equation}
since the conservation law
$\partial_M(\sqrt{-G} \, \overline{\Psi}\hat{\gamma}^M\Psi)=0$ yields the
probability density, where we set
$\overline{\Psi}=\Psi^\dagger{\gamma}^1$.
Although further discussion may be needed on the choice of the normalization,
it is convenient to define $\mathsf{\Psi}\equiv |2U|^{1/4}a^{3/4}\Psi$ so that
\begin{equation}
\mbox{Probability
density}\propto\|\mathsf{\Psi}\|^2\,.
\end{equation}

Now, we compare the solution (\ref{DS1}) of the Dirac-type equation (\ref{DE1})
with the solution (\ref{KGS1}) of the Klein-Gordon-type equation (\ref{KGE1}) in
Model 1 with $K>0$. Asymptotics of the modified Bessel function of the second
kind with complex order are known to be \cite{Bateman}
\begin{equation}
K_{i\frac{\nu}{2}}(y/2)\sim\sqrt{4\pi}e^{-\frac{\nu\pi}{4}}
(\nu^2-y^2)^{-\frac{1}{4}}
\sin\left(\frac{\pi}{4}-\frac{1}{2}\sqrt{\nu^2-y^2}+\frac{\nu}{2}
\cosh^{-1}\frac{\nu}{y}\right)\,,
\end{equation}
and \cite{Tseng}
\begin{eqnarray}
K_{i\frac{\nu}{2}\pm\frac{1}{2}}(y/2)&\sim&\sqrt{4\pi}e^{-\frac{\nu\pi}{4}\pm
i\frac{\pi}{4}}(\nu^2-y^2)^{-\frac{1}{4}}\Biggl[\sqrt{\frac{\nu+y}{2y}}
\sin\left(\frac{\pi}{4}-\frac{1}{2}\sqrt{\nu^2-y^2}+\frac{\nu}{2}
\cosh^{-1}\frac{\nu}{y}\right)\nonumber \\
& &\quad\qquad\qquad\qquad\mp i\sqrt{\frac{\nu-y}{2y}}
\cos\left(\frac{\pi}{4}-\frac{1}{2}\sqrt{\nu^2-y^2}+\frac{\nu}{2}
\cosh^{-1}\frac{\nu}{y}\right)\Biggr]\,,
\end{eqnarray}
as $y\rightarrow\infty$. These functions have a similar oscillatory behavior, 
%R
up to a slowly changing phase shift.

According to the analysis of Kiefer \cite{Kiefer2}, an arranged Gaussian wave
packet tracing a common classical path 
\begin{equation}
a^2=\frac{\bar{\nu}}{\sqrt{K}|p|\cosh 2(\phi-\phi_0)}
\end{equation}
can be constructed with the amplitude ${\cal
A}(\nu)$ whose center $\nu=\bar{\nu}$ takes a relatively large value in both
cases of equations for Model 1, such as %RR
\begin{equation}
{\cal A}(\nu)\propto \frac{1}{(\sqrt{\pi}b)^{1/2}}e^{-(\nu-\bar{\nu})^2/(2b^2)}\,,
\end{equation}
where $b$ represents the width of the wave packet.
The similarity in asymptotics of
$K_{i\frac{\nu}{2}}$ and
$K_{i\frac{\nu}{2}\pm\frac{1}{2}}$ also helps to construct wave packet solutions
for the Dirac-type equation (\ref{LE2}) as well as the Klein-Gordon-type equation
(\ref{LEKG2}) \cite{ALNW} for Model 2. %RR
Note that because the factor $\sqrt{-G}\sqrt{|G^{11}|}$ is proportional to the
argument of the modified Bessel function in both models, the normalized Dirac-type
wave function $\mathsf{\Psi}$ is expressed as a superposition of
$\sqrt{y}K_{i\frac{\nu}{2}\pm\frac{1}{2}}(y)$.

The different behaviors of the functions $K_{i\frac{\nu}{2}}(y/2)$ and
$\sqrt{y/8}K_{i\frac{\nu}{2}\pm\frac{1}{2}}(y/2)$ seems to be especially around
$y=0$. Therefore, the difference between wave-packet solutions of equations of
Klein-Gordon-type (constructed from of
$\psi_{\nu p}$) and Dirac-type (constructed from
$\mathsf{\Psi}_{\pm,\nu p}\equiv|2Ua^3|^{1/4}\Psi_{\pm,\nu p}$) is expected to be
found in the region of small scale factors. % in Model 1.

%R
In this section, we have implicitly accepted the condition $p=$const.
As stated in Sec.~\ref{sec2}, we can consider $\chi$ as a real coordinate
and can also take the wave function which has nontrivial dependence on $\chi$.
It will be, however, the subject to future research. 
%R

%%%%%%%%%%%%%%%%%%%%%%%%%%%%%%%%%%%%%%%%%%%%%%%%%%%%%%%%%%%%%%%%%%%%%%%%%%%
%%%%%%%%%%%%%%%%%%%%%%%%%%%%%%%%%%%%%%%%%%%%%%%%%%%%%%%%%%%%%%%%%%%%%%%%%%%
%%%%%%%%%%%%%%%%%%%%%%%%%%%%%%%%%%%%%%%%%%%%%%%%%%%%%%%%%%%%%%%%%%%%%%%%%%%
\section{Discussion and outlook}
\label{conclusion}
%%%%%%%%%%%%%%%%%%%%%%%%%%%%%%%%%%%%%%%%%%%%%%%%%%%%%%%%%%%%%%%%%%%%%%%%%%%
%%%%%%%%%%%%%%%%%%%%%%%%%%%%%%%%%%%%%%%%%%%%%%%%%%%%%%%%%%%%%%%%%%%%%%%%%%%
%%%%%%%%%%%%%%%%%%%%%%%%%%%%%%%%%%%%%%%%%%%%%%%%%%%%%%%%%%%%%%%%%%%%%%%%%%%

In this paper, the WDW equation has been reformulated as a partial 
differential equation with the Laplacian defined in a minisuperspace extended by
the Eisenhart-Duval lift. We have also obtained the wave function as an exact
fundamental solution to the Dirac equation in the extended minisuperspace of
specific models. We should emphasize that few papers have evaluated the concrete
form of the solutions so far. Further research is needed for general
cosmological models. It is important to study the geometrical property of the
extended minisuperspace and the possibility of geometrical classification on
whether the model can be solved or not. We wish to properly understand the
handling of additional degrees of freedom and the degree of conformal
transformation (and its possible extended symmetry) from various standpoints, as
well as the straightforward generalization to the quantum cosmology of modified
gravities. In any case, the present framework based on the extended
minisuperspace creates many new research agendas.

Here we consider general dimensional cases with conformal coupling in the WDW
equation. As a special case, we can imagine a case that the extended
minisuperspace is conformally flat. Provided that the extended minisuperspace is
conformally flat, the  curvature
${\cal R}$ will be zero if the appropriate conformal transformation, or gauge, is
chosen. Therefore, in this case, the WDW equation is described as a normal Laplace
equation. 

The chosen gauge (\ref{gauge}) at the classical level is a gauge that enables
the simple interpretation when the expanded minisuperspace is
three dimensional and conformally flat. If the extended minisuperspace
is three dimensional, the necessary and sufficient condition for
conformally flatness is that the Cotton tensor is zero \cite{Cotton}. Under the
gauge choice (\ref{gauge}), if the minisuperspace is conformally flat it can be
shown that the (Lorentzian) subspace stretched by variables $a$ and $\phi$ is a
space with constant scalar curvature \cite{GJ1,Jackiw}. Therefore, the scalar
curvature
$r$ of the three-dimensional minisuperspace is a constant. Since the gauge
$G_{\chi\chi}=1$ is chosen, the equation becomes the WDW equation in which the
coefficient before the potential term is $p^2+\xi r$, instead of $p^2$ as in our
previous analysis. In our two models examined in the present paper, the value of
$r$ happened to be zero. Although we have not yet found an example of a model with
nonzero
$r$, the model with three-dimensional extended minisuperspace of conformally
constant curvature belongs to an interesting class of quantum cosmological models.

Another example of a flat three-dimensional extended minisuperspace
with the gauge $G_{\chi\chi}=1$ is
the system with a general nonminimally \cite{Kiefer4} (including conformal
\cite{Hawking,HP2,Kiefer3,GS,Pedram}) coupled scalar field without any
scalar potential term (including the cosmological term) in the spacetime with
nonvanishing spatial curvature.%
\footnote{The extended minisuperspaces for the Kaluza-Klein theory with a
cosmological constant
\cite{Wudka1,Wudka2}, the scalar-tensor theory with a
cosmological constant \cite{Lidsey}, and the Kantowski-Sachs model with a
cosmological constant \cite{Conradi}, which are all known to be related to the
similar model, are also flat.}
 However,
in general cases, the WDW equation may be solved only
when the variable is far from the original variable, which may rather make the
analysis difficult. Generally speaking, including this example, solvability of 
models obtained by the Eisenhart-Duval lift is technically another important
problem, and we would like to classify the cases skillfully in future work.

It should be pointed out that, since the equations obtained in the present paper
are conventional forms of the Laplacian operator and the Dirac operator
which appear in field theories, it is natural to bring them to the third
quantization \cite{SC,CM,McGuigan1,McGuigan2,HM,OEF1,OEF2,Perez} with and without 
introducing the non-linear term of the wave function. We can speculatively imagine
some nonlinear Schr\"odinger equations or the Lane-Emden type equations in the
(extended) minisuperspace. In the context of the third quantization, it is also
interesting to speculate that the global structure of the extended minisuperspace,
such as the compactness of the manifold, will affect the quantum dynamics of the
Universe.

On the other hand, we would like to consider the formulation using the
Eisenhart-Duval metric with more degrees of freedom
\cite{Eisenhart,CPC,Pettini,Cariglia} in quantum cosmology. In addition, although
we have focused on the discussion in the minisuperspace so far, we hope that we
can proceed with the Eisenhart-Duval lift in more basic quantum field theory of
gravity%
\footnote{The geodesics and the quantum Hamiltonian in the
infinite-dimensional superspace have been investigated \cite{DeWitt,FM}.}
 with reference to previous research
\cite{FKP1,FKP2}.

%%%%%%%%%%%%%%%%%%%%%%%%%%%%%%%%%%%%%%%%%%%%%%%%%%%%%%%%%%%%%%%%%
%%%%%%%%%%%%%%%%%%%%%%%%%%%%%%%%%%%%%%%%%%%%%%%%%%%%%%%%%%%%%%%%%
%\appendix
%%%%%%%%%%%%%%%%%%%%%%%%%%%%%%%%%%%%%%%%%%%%%%%%%%%%%%%%%%%%%%%%%

%%%%%%%%%%%%%%%%%%%%%%%%%%%%%%%%%%%%%%%%%%%%%%%%%%%%%%%%%%%%%%%%%%%%%%%%%%%
%%%%%%%%%%%%%%%%%%%%%%%%%%%%%%%%%%%%%%%%%%%%%%%%%%%%%%%%%%%%%%%%%%%%%%%%%%%
%%%%%%%%%%%%%%%%%%%%%%%%%%%%%%%%%%%%%%%%%%%%%%%%%%%%%%%%%%%%%%%%%%%%%%%%%%%
%\section{}\label{AA}
%%%%%%%%%%%%%%%%%%%%%%%%%%%%%%%%%%%%%%%%%%%%%%%%%%%%%%%%%%%%%%%%%%%%%%%%%%%
%%%%%%%%%%%%%%%%%%%%%%%%%%%%%%%%%%%%%%%%%%%%%%%%%%%%%%%%%%%%%%%%%%%%%%%%%%%
%%%%%%%%%%%%%%%%%%%%%%%%%%%%%%%%%%%%%%%%%%%%%%%%%%%%%%%%%%%%%%%%%%%%%%%%%%%

%%%%%%%%%%%%%%%%%%%%%%%%%%%%%%%%%%%%%%%%%%%%%%%%%%%%%%%%%%%%%%%%%%%%%%%%%%%
\acknowledgments
%%%%%%%%%%%%%%%%%%%%%%%%%%%%%%%%%%%%%%%%%%%%%%%%%%%%%%%%%%%%%%%%%%%%%%%%%%%
%Acknowledgements
%%%%%%%%%%%%%%%%%%%%%%%%%%%%%%%%%%%%%%%%%%%%%%%%%%%%%%%%%%%%%%%%%%%%%%%%%%%
%\begin{acknowledgments}
The authors are grateful to P.~Horv\'athy for important information and
references.
The authors thank A.~Paliathanasis for a reference and information on an
interesting recent study \cite{ZP}.
The authors also thank J.~C.~Feng for old and new references.
%the organizers of JGRG21, where our
%partial result %({\tt [arXiv:10mm.xxxx]}) 
%was presented. %for elucidating comments.
%This study is supported in part by the Grant-in-Aid of Nikaido Research 
%Fund.
%\end{acknowledgments}
%%%%%%%%%%%%%%%%%%%%%%%%%%%%%%%%%%%%%%%%%%%%%%%%%%%%%%%%%%%%%%%%%%%%%%%%%%%

%%%%%%%%%%%%%%%%%%%%%%%%%%%%%%%%%%%%%%%%%
%%%%%%%%%%%%%%%%%%%%%%%%%%%%%%%%%%%%%%%%%
%%%
%%%   References
%%%
%%%%%%%%%%%%%%%%%%%%%%%%%%%%%%%%%%%%%%%%%
%%%%%%%%%%%%%%%%%%%%%%%%%%%%%%%%%%%%%%%%%
%%%%%%%%%%%%%%%%%%%%%%%%%%%%%%%%%%%%%%%%%%%%%%%%%%%%%%%%%%%%%%%%%%%%%%%%%%%
%thebibliography
%%%%%%%%%%%%%%%%%%%%%%%%%%%%%%%%%%%%%%%%%%%%%%%%%%%%%%%%%%%%%%%%%%%%%%%%%%%
%\bibliographystyle{apsrev}
\bibliographystyle{apsrev4-1}
%\bibliography{}

%%%%%%%%%%%%%%%%%%%%%%%%%%%%%%%%%%%%%%%%%%%%%%%%%%%%%%%%%%%%%%%%%%%%%%%%%%%

%%%%%%%%%%%%%%%%%%%%%%%%%%%%%%%%%%%%%%%%%%%%%%%%%%%%%%%%%%%%%%%%%%%%%%%%%%%
%%%%%%%%%%%%%%%%%%%%%%%%%%%%%%%%%%%%%%%%%%%%%%%%%%%%%%%%%%%%%%%%%%%%%%%%%%
\end{document}